\documentclass[reprint, amsmath, amssymb, pra, superscriptaddress]{revtex4-2}

\usepackage{graphicx}
\usepackage{dcolumn}
\usepackage{bm}
\usepackage{xcolor}
\usepackage{hyperref}

\begin{document}

\title{Periodic cavity state revivals from atomic frequency combs }
\author{Matthias Zens}
\email[matthias.zens@gmail.com]{}
\affiliation{Institute for Theoretical Physics, Vienna University of Technology (TU Wien), Wiedner Hauptstra\ss e 8-10/136, A--1040 Vienna, Austria, European Union}

\author{Dmitry O.\ Krimer}
\affiliation{Institute for Theoretical Physics, Vienna University of Technology (TU Wien), Wiedner Hauptstra\ss e 8-10/136, A--1040 Vienna, Austria, European Union}
\author{Himadri S.\ Dhar}
\affiliation{Department of Physics, Indian Institute of Technology Bombay, Powai, Mumbai 400076, India}
\author{Stefan Rotter}
\affiliation{Institute for Theoretical Physics, Vienna University of Technology (TU Wien), Wiedner Hauptstra\ss e 8-10/136, A--1040 Vienna, Austria, European Union}

\date{\today}

\begin{abstract}
Spin ensembles with a comb-shaped spectrum have shown exciting properties as efficient quantum memories. Here, we present a rigorous theoretical study of such atomic frequency combs in the strong coupling limit of cavity QED, based on a full quantum treatment using tensor-network methods. Our results demonstrate that arbitrary multi-photon states in the cavity are almost perfectly absorbed by the spin ensemble and re-emitted as parity-flipped states at periodic time intervals. Fidelity values near unity are achieved in these revived states by compensating for energy shifts induced by the strong spin-cavity coupling through adjustments of individual coupling values of the teeth in the atomic frequency comb.
\end{abstract}

\maketitle

\section{Introduction}
Over the years, frequency combs have become an invaluable tool with technological applications in a variety of different fields including high-precision metrology, spectroscopy, attosecond science, and optical communication \cite{Fortier2019,Picque2019}. Optical frequency combs have thereby also entered the quantum domain with exciting prospects for quantum computation and for generating non-classical states of light \cite{Reimer2016,Kues2019}. More recently, frequency combs based on atomic energy levels have been in focus for the design of enhanced quantum memories with long storage times \cite{Laplane2016,Holzapfel2020}, on-demand readout \cite{Jobez2014,Gerasimov2014,Laplane2016,Holzapfel2020}, and large multi-mode capacity \cite{DeRiedmatten2008,Sinclair2014,Jobez2016}. Several groups have demonstrated the storage of pulses containing both single and entangled photons in atomic frequency combs \cite{Clausen2011,Saglamyurek2011,Maring2017, Davidson2020}. 

The efficiency of such quantum memories has been further improved by coupling the atomic frequency comb to a resonant cavity \cite{Afzelius2010,Moiseev2010}.
So far, however, frequency combs have been mostly considered for impedance matched cavities \cite{Afzelius2010,Moiseev2010,Sabooni2013,Zhang2015,Maring2017, Jobez2014} leaving the strong-coupling regime largely unexplored. Furthermore, most studies have focused on the information stored in the amplitude and the relative phase of the incoming pulses \cite{Staudt2007}, while the larger quantum mechanical phase space of the cavity has not received much attention. Importantly, this infinite dimensional Hilbert space can encode and process quantum information in various forms ranging from simple Fock states \cite{Varcoe2000} to error-correcting Schrödinger cat codes \cite{Leghtas2013,Vlastakis2013,Mirrahimi2014,Gertler2020}, binomial codes \cite{Michael2016,Hu2019}, or Gottesman-Kitaev-Preskill codes \cite{Kitaev2001,Campagne2020}, which promise fault-tolerant bosonic quantum computing.   

In this paper we present a theoretical study on the dynamics of arbitrary multi-photon states inside a cavity strongly coupled to an atomic frequency comb. In particular, we show that the strong coupling to the atomic frequency comb leads to a periodic absorption and re-emission of the initial cavity state, equivalent to compressing the cavity's time evolution to stroboscopic revivals. Between these revivals, the cavity state is transferred into the atomic ensemble such that the state's overall lifetime can even exceed the limit imposed by the cavity loss. More specifically, we show that the periodic absorption and re-emission process by the atomic frequency comb acts as a parity transformation on the cavity state---an operation that could be useful in modern bosonic quantum error correcting codes, where the photon number parity plays an important role \cite{Kitaev2001, Campagne2020,Mirrahimi2016,Gertler2020}.

A crucial ingredient to arrive at these results is the insight that the normal-mode splitting in the eigenvalue spectrum of the strongly coupled spin-cavity system distorts the otherwise equidistantly spaced atomic comb structure \cite{Krimer2016}. Here we demonstrate how an equidistant structure in the eigenvalue spectrum can be restored by judiciously adjusting the individual coupling strengths of the comb's teeth. In this way we take full advantage of both the regular frequency spacing of the comb and the efficient information transfer to and from the cavity enabled by the strong-coupling regime \cite{Kubo2010,SaezBlazquez2018,Debnath2020,Weichselbaumer2020,Eisenach2021}. As a result, we obtain a long-lasting train of periodic revivals of the multi-photon cavity state with very high fidelity and minimum losses.

\section{Theoretical model}
Our starting point is the Tavis-Cummings Hamiltonian \cite{Tavis1968}, which models an ensemble of $N$ two-level atoms or spins strongly coupled to a single-mode cavity (using the dipole and rotating wave approximations, $\hbar = 1$),
\begin{equation} 
\mathcal{H} = \omega_{c}~ \hat{a}^\dag \hat{a} +\frac{1}{2} \sum_{k=1}^{N} \omega_{k} ~ \sigma^z_k  + i\sum_{k=1}^{N} g_k(\sigma^+_k \hat{a} - \sigma^-_k \hat{a}^\dag).
\label{Ham}
\end{equation} 
Here, $\omega_c$ is the resonance frequency of the cavity field with the creation and annihilation operators $\hat{a}^\dag$ and $\hat{a}$. Furthermore, $\omega_{k}$ and $g_k$ are the transition frequency and coupling strength for the $k$-th spin and $\sigma^z_k$, $\sigma^+_k$ and $\sigma^-_k$ are the spin-{\footnotesize 1/2} Pauli operators. In general, there are losses in the cavity and spins, which makes it an open system. The dynamics using a Markov-approximation is then given by the Lindblad equation \cite{BRE02}, 
\begin{eqnarray} 
\frac{d\rho}{dt} &=&\mathcal{L}[\rho] = -i[\mathcal{H},\rho]+ \frac{\kappa}{2} ( 2\hat{a} \rho {\hat{a}}^\dag -\hat{a}^\dag \hat{a} \rho-\rho\hat{a}^\dag \hat{a})\nonumber\\&&+ \sum_k \frac{\gamma_h}{2} (2\sigma^-_k\rho\sigma^+_k-\sigma^+_k \sigma^-_k \rho-\rho\sigma^+_k \sigma^-_k)\nonumber\\&&+ \sum_k \gamma_p (\sigma^z_k\rho\sigma^z_k-\rho),
\label{Lind}
\end{eqnarray} 
where $\kappa$ {\color{black} are the cavity losses and $\gamma_{h/p}$ are the radiative/non-radiative losses of the $k$-th spin}. {\color{black} Note that the strong-coupling regime requires some caution in the choice of the master equation \cite{Scala_2007}, and an examination of deviations from Eq.\,\eqref{Lind} in our system could be the scope of future work.} 

We use a time-adaptive variational renormalization group method \cite{Dhar2018} to unravel the full quantum dynamics of a cavity wavefunction strongly coupled to an ensemble of up to one hundred spins. Our numerical procedure relies on the efficient mapping of the extremely large Hilbert space of the cavity-ensemble system, which grows exponentially with the number of spins, to a reduced vector space of computationally tractable size. In particular, we first vectorize the system's density matrix $|\rho\rangle=\textbf{vec}(\rho)$ such that
\begin{align} 
|\rho\rangle = \sum_{k_1,\dots,k_N = 1}^{n_s^2}\sum_{k_c = 1}^{n_c^2} p_{k_1,\dots,k_N,k_c} |k_1\rangle\otimes\dots |k_N\rangle\otimes|k_c\rangle,
\end{align} 
where, $|k_i\rangle$, and $|k_c\rangle$ are the $i$-th spin and Fock superket basis, respectively.
Here, $n_c$ and $n_s$ are the dimensions of the cavity and of the spin operators in the original Hilbert space. Within the vectorized superoperator space the Lindblad equation can be written as ${d|\rho\rangle}/{dt} =\tilde{\mathcal{L}}|\rho\rangle$, where
\begin{equation}
\tilde{\mathcal{L}} = -i(\mathcal{H} \otimes \mathbb{I} - \mathbb{I} \otimes \mathcal{H}^T) +  \kappa  \tilde{\mathcal{L}}_{\hat{a}}+  \sum\mathop{}_{\mkern-5mu k} \gamma_{h,p}  \tilde{\mathcal{L}}_{\sigma^{-,z}_k},
\label{Lind_vec}
\end{equation}
with $\tilde{\mathcal{L}}_{\hat{x}} =  \hat{x} \otimes {\hat{x}}^* - \frac{1}{2}\hat{x}^\dag \hat{x} \otimes \mathbb{I} - \frac{1}{2}\mathbb{I} \otimes \hat{x}^T\hat{x}^*$. This superoperator formalism and the absence of direct dipole-dipole interactions allows us to treat the Lindbladian dynamics of the open quantum system in terms of a variational renormalization group method similar to a central body problem \cite{Stanek2013}. In our case, the cavity acts as the central object, which mediates the interactions between the individual spins. The  superket of the central cavity is always stored exactly, while the spin ensemble is numerically renormalized and truncated at each step in a time-adapative manner \cite{Dhar2018}, similar to a time-evolving block decimation (TEBD) or a time-dependent density matrix renormalization group (\textit{t}-DMRG) method \cite{Vidal2003,Vidal2004,Daley2004,White2004}.

\begin{figure}[ht]
\includegraphics[angle=0,angle=0,width=1.00\columnwidth]{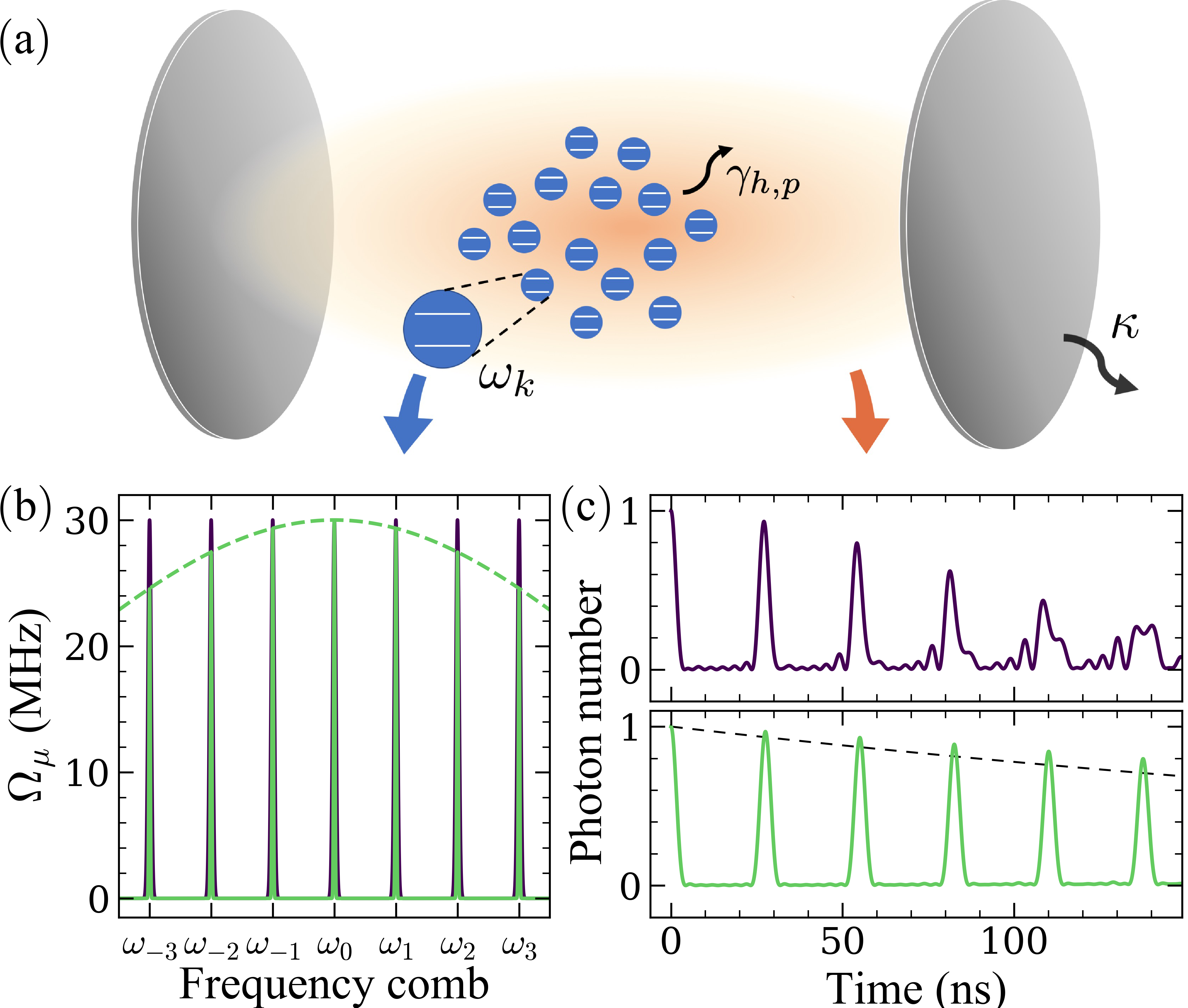}
\vspace*{-0.6cm}
\caption{(a) An ensemble of spins (blue) whose frequencies $\omega_k$ form a frequency comb structure are strongly coupled to a single-mode cavity field (orange). The cavity and spin losses are given by $\kappa$ and $\gamma_{h,p}$. (b) Comb-shaped distribution of the collective coupling strengths $\Omega_\mu$. Here we compare a uniform comb structure, where each frequency cluster (indicated by a delta-like peak) couples to the cavity with the same coupling strength $\Omega_\mu/2\pi=30$ MHz (purple), with a spectrally engineered comb structure (light green), where $\Omega_\mu$ follows a Gaussian envelope (dashed line) $\Omega_\mu = \Omega_0\exp\left[-(\omega_c-\omega_\mu)^2/2\lambda^2\right]$, with $\Omega_0/2\pi=30$ MHz and $\lambda/2\pi=0.19$ GHz. (c)  Cavity photon number $\langle a^\dag a\rangle$ as a function of time for the modified (bottom) and uniform (top) comb structure. The spin ensemble in both cases is initially unexcited and the cavity is prepared in a coherent state $|\alpha\rangle$ of amplitude $\alpha=1$. The black dashed line corresponds to the bare cavity decay proportional to $\exp(-\kappa t)$, with $\kappa/2\pi=0.4$ MHz.}
\label{Fig1}
\end{figure}

\section{Atomic frequency comb}

\begin{figure*}[ht]
\includegraphics[angle=0,angle=0,width=2.00\columnwidth]{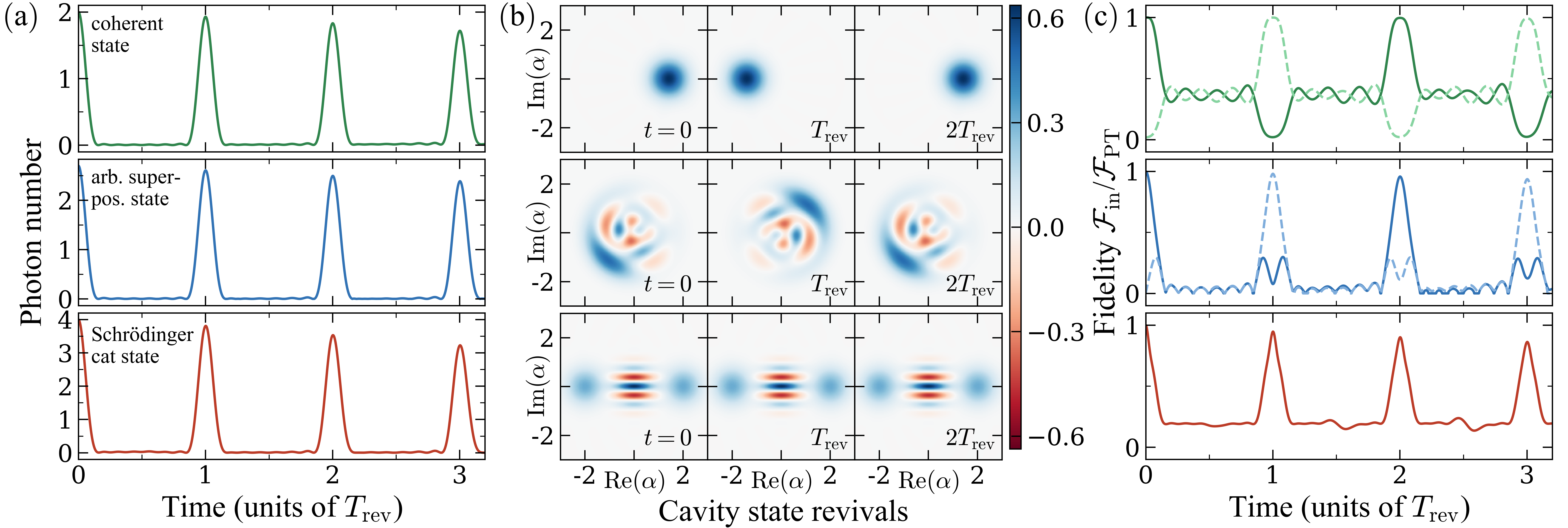}
\vspace*{-0.2cm}
\caption{(a) Cavity photon number $\langle a^\dag a\rangle$ as a function of time for three different initial states $|\psi_\mathrm{cav}^\mathrm{in}\rangle$ of the cavity: (i) $|\alpha\rangle$, (ii) $|\psi_\mathrm{sup}\rangle$, and (iii) $|\psi_\mathrm{cat}\rangle$ (top to bottom) as specified in the main text. (b) Wigner functions of the cavity states (i)-(iii) at $t=0$, at the first revival ($t=T_\mathrm{rev}$) and the second one  ($t=2T_\mathrm{rev}$). Note that after one absorption and re-emission period $T_\mathrm{rev}$, the Wigner function is point-reflected through the origin, which amounts to a parity transformation $\hat{\Pi}=\exp(i\pi\hat{a}^\dag\hat{a})$. (c) Fidelity of the time evolved cavity state $\rho_\mathrm{cav}(t)$ with its initial state from (a) $|\psi_\mathrm{cav}^\mathrm{in}\rangle$ ($\mathcal{F}_\mathrm{in}$, solid line) and with the parity transformed state $\hat{\Pi}|\psi_\mathrm{cav}^\mathrm{in}\rangle$ ($\mathcal{F}_\mathrm{PT}$, dashed line). Here, the coupling distribution $\Omega_\mu$ for all three cases follows the same Gaussian envelope with $\lambda/2\pi=0.19$ GHz. }
\label{Fig2}
\end{figure*}

The salient feature of an atomic frequency comb is the periodic absorption and subsequent re-emission of photons resulting from the comb-shaped frequency distribution of the spin ensemble as sketched in Fig.~\ref{Fig1}. {\color{black} Note that such a frequency comb structure can be prepared using spectral hole burning \cite{Jobez2016,Rubio_2018,Debnath_2019}}. In the present case, the frequency comb is centered around the cavity frequency $\omega_c$, such that the spin frequencies $\omega_\mu=\omega_c+\mu\Delta\omega$ for $\mu=\{-(m-1)/2,...,(m-1)/2\}$ with $m$ (odd) being the number of frequency clusters of the comb. For an ensemble of $N$-spins, each frequency tooth $\omega_\mu$ in the comb corresponds to a sub-ensemble of $N'$ = $N/m$ spins. Accordingly, $\Omega_\mu=\sum_k^{N'} g_{\mu,k}^2 = N'g_\mu^2$ denotes the collective coupling strength of the $\mu$-th sub-ensemble. Throughout this paper we assume $m=7$, $N=70$, $\Delta\omega/2\pi= 40$ MHz, and $\omega_c/2\pi=3$ GHz. If not stated otherwise we use the parameters $\kappa/2\pi=0.4$ MHz, $\gamma_h/2\pi=1$ kHz, and {\color{black}$\gamma_p/2\pi=33$ kHz, which are realistic values for ensembles of nitrogen-vacancy centers coupled to superconducting microwave resonators \cite{Angerer2017}.} 

In the example of Fig.~\ref{Fig1}(c) we assume that the spin ensemble is initially unexcited and the cavity is prepared in a coherent state $|\psi_\mathrm{cav}^\mathrm{in}\rangle=|\alpha\rangle$ of amplitude $\alpha=1$. For the simplest case of a uniform coupling distribution, where each subensemble couples with the same strength $\Omega_\mu/2\pi=30$ MHz, we observe that the cavity excitation is first transferred into the frequency comb and then re-emitted back into the cavity at a well-defined later time. This process then repeats itself periodically, leading to a train of revivals with corresponding cavity photon numbers that drop already significantly during the first five revivals as depicted in Fig.~\ref{Fig1}(c) (top panel). We emphasize that this drastic decrease is not a result of losses in the cavity or in the spin ensemble but rather a consequence of the strong coupling between the cavity mode and the atomic frequency comb, which distorts the comb's equidistant frequency spectrum due to normal-mode splitting. As detailed below, this frequency detuning can, however, be pre-compensated through a customized engineering of the spectral coupling distribution $\Omega_\mu$. Using a Gaussian distribution of standard deviation $\lambda/2\pi=0.19$ GHz for the coupling strengths $\Omega_\mu$, we obtain a long-lived train of revivals in the cavity photon number of which we again show the first five peaks in Fig.~\ref{Fig1}(c) (bottom panel).

A remarkable aspect to be highlighted is the fact that the coupling of the cavity to the frequency comb produces photon numbers in the stroboscopic cavity revivals that may even exceed the limit imposed by the bare cavity decay. A comparison with this exponential decay $\propto\exp(-\kappa t)$, is shown in the bottom panel of Figure~\ref{Fig1}(c). We thus observe that the excitation transfer to the long-lived spin ensemble protects the initial excitation from the cavity losses{\color{black}, which can be associated with the so-called cavity protection effect \cite{Kurucz_2011,Diniz_2011,Putz_2014}}. The important open question we now address is, which quantum states are written into the cavity by the frequency comb during the periodic revivals and how these states are related to the initial cavity wave function.

\section{Cavity state revival and parity transformation}
To answer this question comprehensively, we start again from an unexcited spin ensemble and a cavity prepared \cite{Prep} in three different initial states $|\psi_\mathrm{cav}^\mathrm{in}\rangle$: (i) $|\alpha\rangle$ is a coherent state of amplitude $\alpha=\sqrt{2}$, which closely resembles the situation of a short coherent pulse injected into the cavity. (ii) $|\psi_\mathrm{sup}\rangle=\frac{1}{\mathcal{N}_\mathrm{sup}}\sum_{n=1}^4c_n|n\rangle$ is a superposition of the four lowest Fock states $|n\rangle$ with the coefficients $c_{1-4}$ 
chosen arbitrarily as $5$, $-i\sqrt{15}$, $-(\sqrt{10}-i\sqrt{15})$, and $(5-i\sqrt{10})$ such that no apparent phase relation can be established between neighboring Fock states. (iii) $|\psi\rangle_\mathrm{cat}=\frac{1}{\mathcal{N}_\mathrm{cat}}(|\beta\rangle+|\!-\!\beta\rangle)$ with $\beta=2$ denotes a Schr\"odinger cat state as used in many quantum information processing tasks, including quantum computation \cite{Ralph2003}, quantum teleportation \cite{Enk2001}, and precision measurements \cite{Munro2002}. 

First of all, we see in Fig.~\ref{Fig2}(a) that for all three initial states the cavity photon number $\langle\hat{a}^\dag\hat{a}\rangle$  shows the characteristic periodic revival structure:  The initial cavity photons are absorbed by the spin ensemble within a few nanoseconds, irrespective of the quantum state in which the cavity is initialized. After that, the cavity remains empty until the number of photons is restored at $t=T_\mathrm{rev}\approx 27.5$ ns and the process starts all over again.

The time-adaptive renormalization group method described before now provides us direct access to the full density matrix of the cavity field during the absorption and revival process. 
Using this information, we plot in Fig.~\ref{Fig2}(b) the Wigner function $W(\alpha,\alpha^*,t)=\frac{1}{\pi^2}\int d^2\beta e^{\alpha\beta^*-\alpha^*\beta}\,\mathrm{Tr}\{e^{\beta\hat{a}^\dag-\beta^*\hat{a}}\rho_\mathrm{cav}(t)\}$ of the cavity field  at the initial time $t=0$ and at the first and second revival, $t=T_\mathrm{rev}, 2\,T_\mathrm{rev}$, respectively. We see that at the first revival the Wigner function is point-reflected through the origin as compared to the Wigner function at $t=0$, corresponding to a parity transformation $\hat{\Pi}=\exp(i\pi\hat{a}^\dag\hat{a})$ of the initial cavity state. To confirm this observation, we show in Fig.~\ref{Fig2}(c) the fidelity $\mathcal{F}_\mathrm{in(PT)}=\mathcal{F}\{\rho_\mathrm{cav}(t),\rho_\mathrm{in(PT)}\}$ of the cavity state $\rho_\mathrm{cav}(t)$ with the initial state $\rho_\mathrm{in}=|\psi_\mathrm{cav}^\mathrm{in}\rangle\langle\psi_\mathrm{cav}^\mathrm{in}|$ and with the parity-transformed initial state $\rho_\mathrm{PT}=\hat{\Pi}\rho_\mathrm{in}\hat{\Pi}^\dag$, respectively. For the states $|\alpha\rangle$ and $|\psi_\mathrm{sup}\rangle$,  which are not parity eigenstates, the fidelity with the initial state $\mathcal{F}_\mathrm{in}$ is almost zero at the first revival, while the fidelity with the parity-transformed initial state $\mathcal{F}_\mathrm{PT}$ is close to one. As a consequence, the initial cavity state is restored with very high fidelity only after a period of $2\,T_\mathrm{rev}$. The situation is different for the Schr\"odinger cat state $|\psi_\mathrm{cat}\rangle$, which, being an eigenstate of the parity operator, is restored by the frequency comb already at the first revival $t=T_\mathrm{rev}$, marked by a fidelity $\mathcal{F}_\mathrm{in}=\mathcal{F}_\mathrm{PT}$ equal to one already at that earlier time. 

\section{Spectral engineering of the AFC}
\begin{figure}[htbp]
\includegraphics[angle=0,angle=0,width=1.00\columnwidth]{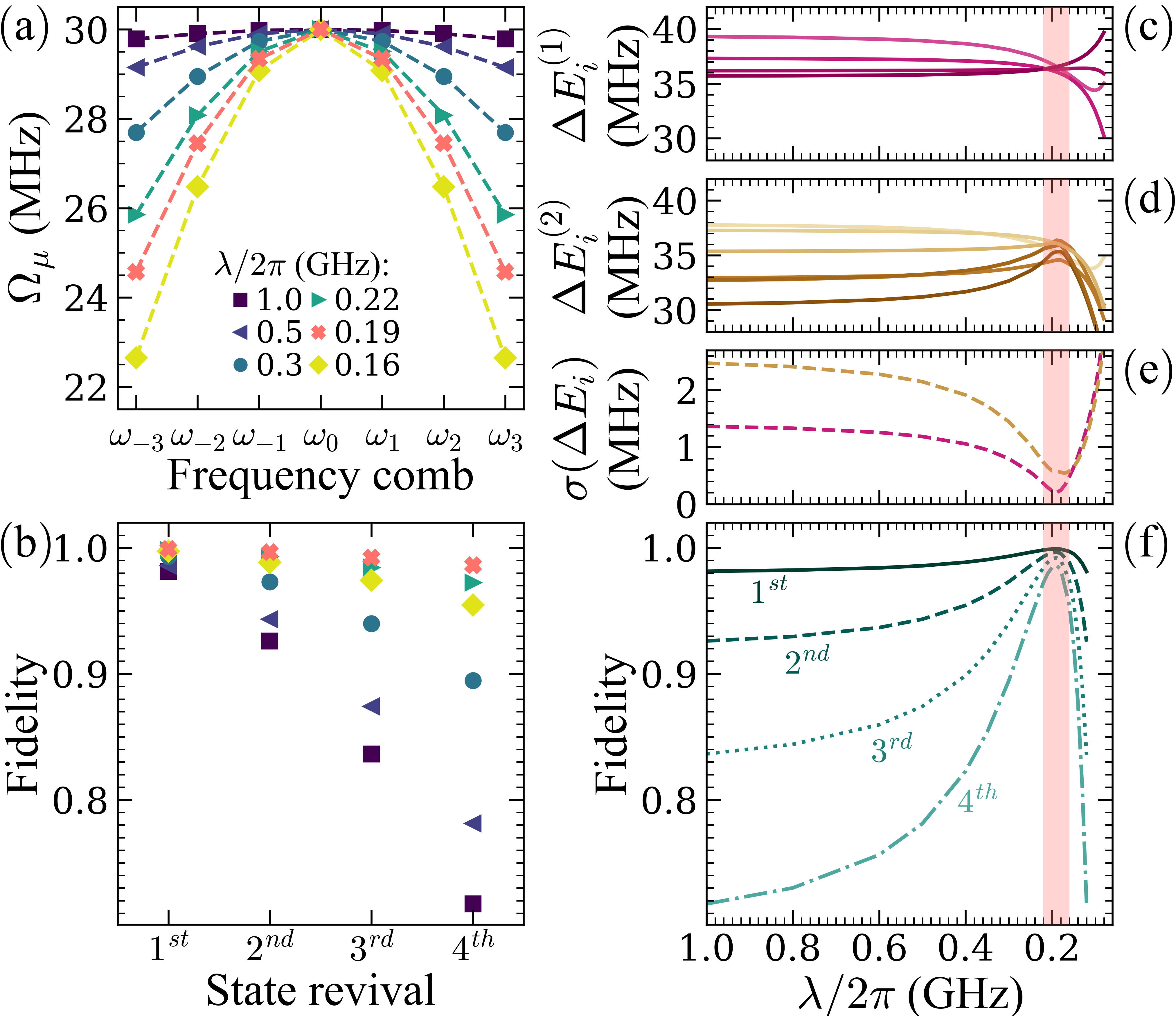}
\vspace*{-0.5cm}
\caption{(a) Distribution of collective coupling strengths $\Omega_\mu$ of each frequency cluster of the atomic frequency comb with $\omega_\mu=\omega_c+\mu\Delta\omega$ for $\mu=\{-3,-2,...,3\}$ and with $\Delta\omega/2\pi=40$ MHz. The couplings follow Gaussian distributions with standard deviations $\lambda/2\pi$ ranging from $1$ GHz to $0.16$ GHz. (b) Fidelity at the first four cavity state revivals for the coupling distributions presented above.  At even revivals the fidelity is calculated for the cavity state function and the initial state $|\psi_\mathrm{cav}^\mathrm{in}\rangle=\frac{1}{\sqrt{2}}(|1\rangle+|2\rangle)$; at odd revivals we calculate the fidelity of the cavity state function and the parity transformed initial state. (c,d) Energy spacings $\Delta E_i^{(1/2)}$ of the one- and two-excitation subspace of the Tavis-Cummings Hamiltonian [Eq.\eqref{Ham}] as a function of the distribution parameter $\lambda$. In the vicinity of $\lambda/2\pi=0.19$ GHz (red shaded areas) the energy levels become almost equidistant resulting in an enhanced performance of the atomic frequency comb. (e) Standard deviation of $\Delta E_i^{(1/2)}$ (purple/yellow) as a function of $\lambda$. (f) Fidelity at the first four cavity state revivals as a function of $\lambda$ showing distinct maxima around $\lambda/2\pi=0.19$ GHz. }
\label{Fig3}
\end{figure}

As indicated in Fig.~\ref{Fig1}, a Gaussian envelope in the distribution of coupling strengths can drastically increase the performance of the atomic frequency comb. The choice of the Gaussian turns out to be sufficient to correct the comb's spectral distortions due to the strong-coupling to the cavity. Moreover, it has the appealing advantage that we can work with the standard deviation $\lambda$ of the Gaussian envelope as the only free parameter, which we tune between the values $\lambda/2\pi=1$ GHz and $\lambda/2\pi=0.1$ GHz as depicted in Fig.~\ref{Fig3}(a) for specific values of $\lambda$. We calculate the corresponding dynamics for the simple initial state $|\psi_\mathrm{cav}^\mathrm{in}\rangle=\frac{1}{\sqrt{2}}(|1\rangle+|2\rangle)$ and evaluate the fidelity $\mathcal{F}_\mathrm{in/PT}$ at the first four revivals. Hereafter we will drop the subscripts in $\mathcal{F_\mathrm{in/PT}}$ implying that we use the former at even revivals and the latter at odd revivals of the cavity state. To exclude effects stemming from the openness of the system, all loss parameters are set to zero such that deviations from $\mathcal{F}=1$ can be attributed exclusively to an imperfect rephasing of the frequency comb. Figure~\ref{Fig3}(b) shows the fidelity $\mathcal{F}$ at the first four revivals for the same selection of coupling distributions as presented in (a).  For $\lambda/2\pi=1$ GHz, which corresponds to an almost uniform coupling, the fidelity stays noticeably below one with $\mathcal{F}=98.1\%$ at the first revival and continues to deteriorate rapidly reaching values ranging from $92.6\%$ to $71.7\%$ for the following three revivals. The fidelity of the revivals drastically increases for decreasing values of $\lambda$ reaching a maximum for $\lambda/2\pi\approx 0.19$ GHz; here the fidelity with the initial state is $98.6\%$ even for the fourth revival. For $\lambda$ decreasing even further, the fidelity of the revivals deteriorates again.

This behavior can best be understood by examining the energy levels of the strongly coupled spin-cavity system \cite{Suppl}. Here, the strong coupling leads to a normal-mode splitting lifting the degeneracy of the cavity mode and the central (resonant) spins. Consequently, the first rung of the energy ladder in the strong coupling regime consists of $m+1$ levels instead of $m$ in the uncoupled case. First, we notice that the mean energy spacing $\overline{\Delta E}_i^{(1)}\approx36.36$ MHz (at $\lambda/2\pi=0.19$ GHz) of these $m+1$ levels is reduced as compared to the uncoupled case $\overline{\Delta E}_i^{(1)}=\Delta\omega/2\pi=40$ MHz. This leads to a revival time of $T_\mathrm{rev}=1/\overline{\Delta E}_i^{(1)}=27.5$ ns, which is in excellent agreement with the value observed in Fig.~\ref{Fig1}. Next, we present in Figure~\ref{Fig3}(c,d) the energy level spacings $\Delta E_i^{(1)}$, $\Delta E_i^{(2)}$ of the one- and two-excitation subspace of the Tavis-Cummings Hamiltonian, Eq.~\eqref{Ham}, as a function of the Gaussian coupling distribution width $\lambda$. The energy shifts induced by the normal-mode splitting are larger for the energy levels close to resonance with the cavity than for off-resonant levels. The energy levels of a uniformly coupled spin-cavity system are therefore no longer equidistant, which inhibits a perfect rephasing of the initial cavity state. Conversely, the Gaussian modification of the coupling distribution introduced above acts as a compensation for the induced energy shifts, as evidenced in Fig.\,\ref{Fig3}(e). Here, we present the standard deviation $\sigma(\Delta E_i^{(1/2)})$ of the energy spacings $\Delta E_i^{(1/2)}$ showing a minimum at $\lambda=0.19$ GHz, which is the same parameter value for which the fidelity depicted in Fig.~\ref{Fig3}(f) shows a maximum. {\color{black} We expect a similar behavior for higher energy levels since the initial states used in our calculations for Fig.~\ref{Fig2} already carry significant multi-photon contributions.} Our findings thus confirm that in the regime of strong coupling, the spectral engineering of the spin ensemble is a viable tool to efficiently preserve the quantum information in the system.

\section{Conclusions}
Our analysis provides the first rigorous and fully quantum mechanical treatment of atomic frequency combs 
in the strong coupling regime of cavity QED. We demonstrate that arbitrary cavity states, ranging from a superposition of low-energy Fock states to macroscopic Schr\"odinger cat states,  can be transferred to a spectrally comb-shaped spin ensemble and retrieved almost perfectly at well-defined later times. The absorption and re-emission by the atomic frequency comb thereby act as a parity transformation on the initial cavity state. Energy shifts induced by the cavity-spin coupling lead to a significant amount of dephasing in the strong coupling regime, but can be pre-compensated by engineering the distribution of coupling strengths in the comb. In this way equidistant energy levels of the coupled spin-cavity system are ensured, resulting in a revival fidelity well above $98.6\%$ for the first four revivals.

\begin{acknowledgments}
 We would like to thank R. Bekenstein and H. Pichler for helpful discussions and acknowledge support by the Austrian Science Fund (FWF) through the Lise Meitner programme, Project No.~M 2022-N27 and the European Commission under Project NHQWAVE No. MSCA-RISE 691209. M.\,Z.\ would like to thank the Institute for Theoretical Atomic, Molecular, and Optical Physics (ITAMP) at Harvard for hospitality and the Austrian Science Fund (FWF) for support through the Doctoral Programme CoQuS (W1210). The computational results presented have been achieved using the Vienna Scientific Cluster (VSC).
\end{acknowledgments}
\bibliographystyle{apsrev4-2}
\bibliography{references.bib}

\end{document}


\title{Periodic cavity state revivals from atomic frequency combs -- Supplemental Material}
\author{Matthias Zens}
\email[matthias.zens@gmail.com]{}
\affiliation{Institute for Theoretical Physics, Vienna University of Technology (TU Wien), Wiedner Hauptstra\ss e 8-10/136, A--1040 Vienna, Austria, EU}
\author{Dmitry O.\ Krimer}
\affiliation{Institute for Theoretical Physics, Vienna University of Technology (TU Wien), Wiedner Hauptstra\ss e 8-10/136, A--1040 Vienna, Austria, EU}
\author{Himadri S.\ Dhar}
\affiliation{Department of Physics, Indian Institute of Technology Bombay, Powai, Mumbai 400076, India}
\author{Stefan Rotter}
\affiliation{Institute for Theoretical Physics, Vienna University of Technology (TU Wien), Wiedner Hauptstra\ss e 8-10/136, A--1040 Vienna, Austria, EU}

\date{\today}

\maketitle

\section{Energy spectrum}
In this supplementary section, we show how the strong spin-cavity coupling distorts the eigenvalue spectrum of the uniformly coupled frequency comb.  Introducing collective spin operators for each subensemble $J_\mu^z=\frac{1}{2}\sum_{k=1}^{N'}\sigma_k^z$ and $J_\mu^\pm=\sum_{k=1}^{N'}\sigma_k^\pm$, the Tavis-Hamiltonian given in Eq.(1) of the main text can be rewritten as  

\begin{equation}
\label{eq:H}
H=\omega_ca^\dagger a + \sum_{\mu}\omega_\mu J_\mu^z + g_\mu \sum_{\mu}(J_\mu^+a+J_\mu^-a^\dagger),
\end{equation}
where $\mu=\{-(m-1)/2,...,(m-1)/2\}$, $m=7$ is the number of subensembles, and $N'=N/m$ is the number of spins in each subensemble. In the following, spins inside the same frequency subensemble are described in the collective spin basis $\ket{J_\mu,m_\mu}=\ket{N'/2,-N'/2+q_\mu}\equiv\ket{q_\mu}$, where $q_\mu$ is the number of excitations in the $\mu$-th subensemble. With the use of
\begin{align}
J_\mu^z\ket{J_\mu,m_\mu}&=m_\mu\ket{J_\mu,m_\mu}=(-N'/2+q_\mu)\ket{q_\mu},
\end{align}
and
\begin{align}
J_\mu^\pm\ket{J_\mu,m_\mu}&=\sqrt{J_\mu(J_\mu+1)-m_\mu(m_\mu\pm1)}\ket{J_\mu,m_\mu}\nonumber\\&=\sqrt{N'/2+N'q_\mu-q_\mu^2\pm(N'/2-q_\mu)}\ket{q_\mu},
\end{align}
we can set up the single- and two-excitation subspace for the coupled spin-cavity system. We write the $1+m$ basis states of the single-excitation subspace of the combined spin-cavity system as $\ket{1_c}\ket{0_\mu}$ and $ \ket{0_c}\ket{1_\mu}$. Here, $\ket{1_c}\ket{0_\mu}$ denotes a state with a single excitation in the cavity part and  no excitations in the spin ensemble, whereas $\ket{0_c}\ket{1_\mu}$ denotes $m$ states where the single excitation is in the $\mu$-th subensemble (with all other subensembles unexcited) and no excitations in the cavity. The action of the Hamiltonian \eqref{eq:H} in this single-excitation basis is the given by
\begin{align}
H\ket{1_c}\ket{0_\mu}&=\left(\omega_c-\frac{N'}{2}\omega_\Sigma\right)\ket{1_c}\ket{0_\mu}+\sum_\mu \Omega_\mu\ket{0_c}\ket{1_\mu},\\  \label{eq4}
H\ket{0_c}\ket{1_\mu}&=\left(\omega_\mu-\frac{N'}{2}\omega_\Sigma\right)\ket{0_c}\ket{1_\mu}+\Omega_\mu\ket{1_c}\ket{0_\mu},
\end{align}
with $\omega_{\Sigma}=\sum_{\mu}\omega_\mu$. 

\begin{figure}[b]
\includegraphics[angle=0,angle=0,width=1.00\columnwidth]{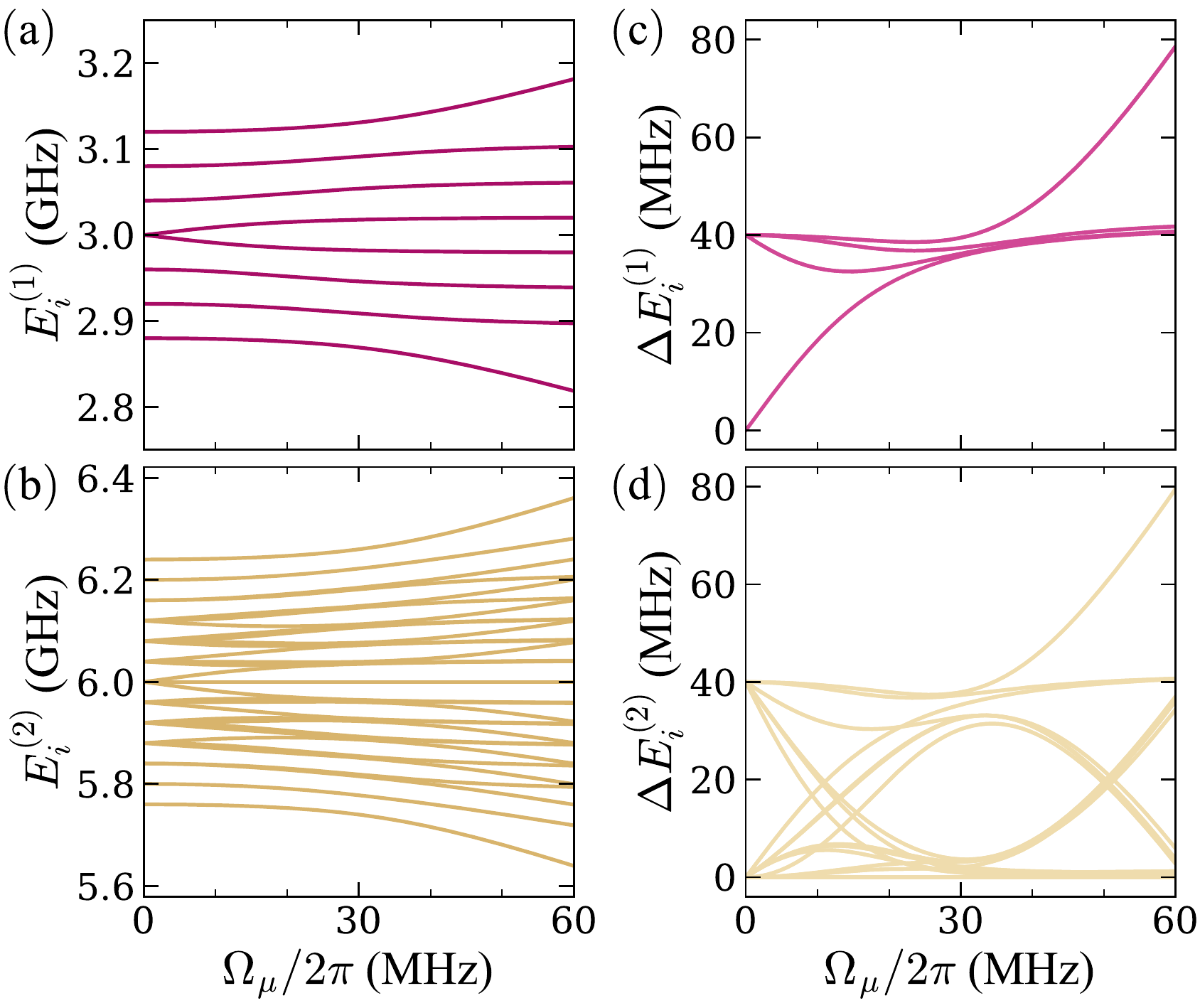}
\vspace*{-0.4cm}
\caption{(a,b) Energy spectrum of the singe- and two-excitation subspace of the Tavis-Cummings Hamiltonian \eqref{eq:H} as a function of the collective coupling strength $\Omega_\mu$. Here we consider the simplest case of an equidistant atomic frequency comb structure with $\Delta\omega/2\pi=40$ MHz uniformly coupled to a single cavity mode. (c,d) Corresponding spacings $\Delta E_i^{(1/2)}$ of neighboring energy levels. The spin-cavity coupling lifts the equidistant energy spectrum of the atomic frequency comb yielding a non-equidistant spectrum for the compound spin-cavity system.}
\label{FigSI1}
\end{figure}

Using the same notation as above, the basis states of the two-excitation subspace can be written as $\ket{2_c}\ket{0_\mu}$, $\ket{1_c}\ket{1_\mu}$, $\ket{0_c}\ket{1_\mu}\ket{1_\nu}$, $\ket{0_c}\ket{2_\mu}$ (in total $1+m+m\;(m-1)/2+m$ basis states). The Hamiltonian \eqref{eq:H} acting on these states gives
\begin{align}
H\ket{2_c}\ket{0_\mu}=&\left(2\omega_c-\frac{N'}{2}\omega_\Sigma\right)\ket{2_c}\ket{0_\mu}\nonumber\\ &+\sqrt{2}\sum_\mu\Omega_\mu\ket{1_c}\ket{1_\mu},
\end{align}

\begin{align}
H\ket{1_c}\ket{1_\mu}=&\left(\omega_c+\omega_\mu-\frac{N'}{2}\omega_\Sigma\right)\ket{1_c}\ket{1_\mu}\nonumber\\ &+\Omega_\mu\sqrt{2}\ket{2_c}\ket{0_\mu}+\sum_{\nu\neq \mu}\Omega_\nu\ket{0_c}\ket{1_\mu}\ket{1_\nu}\nonumber\\ &+\Omega_\mu\sqrt{2-2/N'}\ket{0_c}\ket{2_\mu},
\end{align}

\begin{align}
H\ket{0_c}\ket{1_\mu}\ket{1_\nu}&=\left(\omega_\mu+\omega_\nu-\frac{N'}{2}\omega_\Sigma\right)\ket{0_c}\ket{1_\mu}\ket{1_\nu}\nonumber\\ &+\Omega_\mu\ket{1_c}\ket{1_\nu}+\Omega_\nu\ket{1_c}\ket{1_\mu}\\
H\ket{0_c}\ket{2_\mu}=&\left(2\omega_\mu-\frac{N'}{2}\omega_\Sigma\right)\ket{0_c}\ket{2_\mu}\nonumber\\ &+\Omega_\mu\sqrt{2-2/N'}\ket{1_c}\ket{1_\mu}.\label{eq9}
\end{align}

With the above equations one can set up the Tavis-Cummings Hamiltonian in the single- and two-excitation basis and solve its eigenvalues numerically. The resulting energy spectrum is presented in Fig.~\ref{FigSI1} as a function of the coupling strength $\Omega_\mu$ for the simple case of an equidistant frequency comb with $\Delta\omega/2\pi=40$ MHz and uniform coupling. The strong coupling leads to a normal-mode splitting, which lifts the degeneracy of the cavity mode and the central (resonant) spins. Consequently, the first rung of the energy ladder in the strong coupling regime consists of $m+1$ levels instead of $m$ in the uncoupled case. The coupling between the cavity and spin ensemble thereby shifts the energy levels of the spin-cavity system such that they are no longer equidistant for strong coupling $\Omega_\mu$. In the main text, we show how this drawback of the strong-coupling regime can be overcome by adjusting the individual coupling strengths of the frequency comb.
